\documentclass[aps,prl,superscriptaddress,twocolumn,showpacs]{revtex4-1}
\pdfoutput=1
\usepackage{graphicx,color,amsmath,amsthm}
\usepackage[ugly]{nicefrac}
\begin{document}
\title{Gelation impairs small molecule migration in polymer mixtures}

\author{Biswaroop Mukherjee}
\affiliation{Department of Physics and Astronomy, University of Sheffield, Hounsfield Road, Sheffield S3 7RH, UK.}
\author{Buddhapriya Chakrabarti}
\email{b.chakrabarti@sheffield.ac.uk}
\affiliation{Department of Physics and Astronomy, University of Sheffield, Hounsfield Road, Sheffield S3 7RH, UK.}

\date{\today}

\begin{abstract}
Surface segregation of the low-molecular weight component in a polymeric mixture leads to degradation of industrial formulations. We report a simultaneous phase separation and surface migration phenomena in oligomer-polymer and oligomer-gel systems following a temperature quench. We compute equilibrium and time varying migrant density profiles and wetting layer thickness using coarse grained molecular dynamics and mesoscale hydrodynamics  simulations to demonstrate that surface migration in oligomer-gel systems is significantly reduced due to network elasticity. Further, phase separation processes are significantly slowed in gels, modifying the Lifshitz-Slyozov-Wagner (LSW) law $\ell(\tau) \sim \tau^{1/3}$. Our work allows for rational design of polymer/gel-oligomer mixtures with predictable surface segregation characteristics. 

%Surface segregation of the low molecular weight component of a polymeric mixture is a ubiquitous phenomenon that leads to degradation of industrial formulations. We report a simultaneous phase separation and surface migration phenomena in oligomer-polymer ($OP$) and oligomer-gel ($OG$) systems following a temperature quench that induces demixing of components. We compute equilibrium and time varying migrant (oligomer) density profiles and wetting layer thickness in these systems using coarse grained molecular dynamics (CGMD) and mesoscale hydrodynamics (MH) simulations. Such multiscale methods quantitatively describe the phenomena over a wide range of length and time scales. We show that surface migration in gel-oligomer systems is significantly reduced on account of network elasticity. Further, the phase separation processes are significantly slowed in gels leading to the modification of the well known Lifshitz-Slyozov-Wagner (LSW) law $\ell(\tau) \sim \tau^{1/3}$. Our work allows for rational design of polymer/gel-oligomer mixtures with predictable surface segregation characteristics that can be compared against experiments. 
\end{abstract}

\maketitle

\textit{Introduction:} Complex mixtures of soft materials, used in industrial formulations, are often plagued by the migration of the small molecular weight component to the interface open to atmosphere\cite{b:r.a.l.jones.1999}. Such ``surface segregation'' of the active ingredients in a formulation leads to loss of function in a variety of our daily products\cite{p:lonchampt.ejlst.v106.p241.y2004,p:rasco.comprevfoodsci.v12.p523.y2013} \textit{e.g.} adhesive loss in feminine and hygiene care products, flaking behaviour of paints, and blooming of fat in chocolate. The basic phenomenology of surface segregation/wetting can be understood in a model binary polymer mixture of different molecular weights and a surface exposed to atmosphere. The surface composition of this mixture is determined by the relative surface energies of individual components. A loss of entropy and gain in surface energy of a molecule dictates the equilibrium surface fraction. 

For well-mixed systems having a free surface, the segregation profile with oligomer concentration monotonically decreasing as a function of depth $(\approx \exp(-z/\xi))$ is observed. In contrast a macroscopic ($\ell_{w} \approx 100-200$ nm) wetting layer forms for systems for which the bulk thermodynamic phase is de-mixed. The classic experiments demonstrating surface directed spinodal decomposition (SDSD) were performed on an unstable polymer mixture of PEP and dPEP having a free surface, which preferentially attracts dPEP \cite{p:r.a.l.jones.prl.v66.p1326.y1991,p:r.a.l.jones.pre.v47.p1437.y1993,p:r.a.l.jones.polymer.v35.p2160.y1994}. Mean field (MFT) and self-consistent field theories (SCFT) that combine the bulk thermodynamics of polymer mixtures with that of a surface expressed in terms of phenomenological free energy funtionals have been employed to compute the surface migrant fraction and wetting layer thickness for well mixed and demixed systems with moderate success. These theories however do not describe how the migrant concentration profiles and wetting layers evolve as a function of time. 

In an earlier study\cite{p:self.prl.v116.p208301.y2016} we showed that increasing the bulk modulus of a gel-oligomer mixture causes a dramatic reduction in the surface fraction of migrant molecules. The wetting transition observed for demixed systems can also be avoided. This study was based on a mean field analysis of a phenomenological free energy functional. We augment this study with CGMD simulations which gives a more accurate representation of the physical situation particularly near a phase transition. 
\begin{figure}[t]
\includegraphics[width=8.5cm]{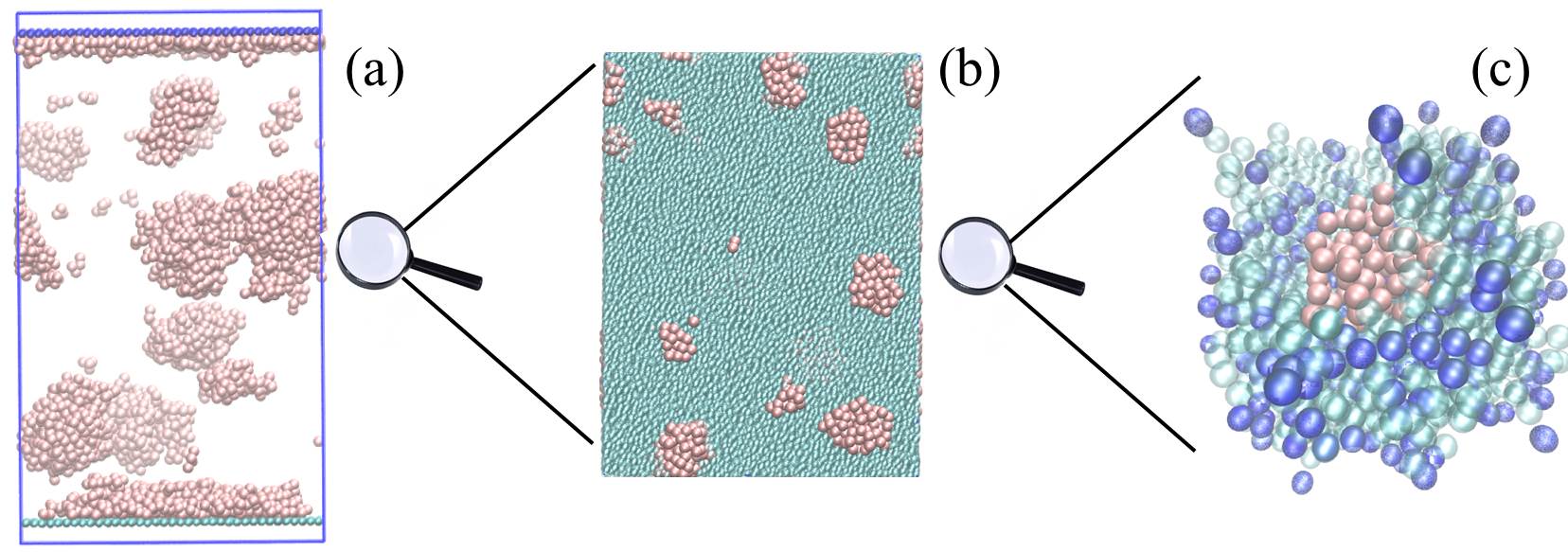} 
\caption{Configuration snapshots of a gel-oligomer system (CGMD) undergoing phase separation at different scales of resolution. Panel (a) shows the simulation box with oligomers droplets (pink), while (b) shows a magnified region of oligomer droplets (pink) in a gel-matrix (green). Panel (c) shows an oligomer droplet (pink beads) trapped within a mesh of an end-linked gel formed by polymers (green beads) with permanently stuck end groups (blue beads).\label{fig01}}
\end{figure}

In this letter we report the kinetics of surface migration of small molecules (oligomers) in a {\it (a)} oligomer-polymer ($OP$)and {\it (b)} oligomer-gel ($OG$) mixture undergoing phase separation following an instantaneous temperature quench that renders the mixed phase unstable. The surface free energy difference preferentially attracts oligomers. We {\it (i) compute dynamic surface concentration profiles of oligomers, (ii) quantify the difference in bulk coarsening phenomena as a function of depth of quench $\Delta T$ and gel bulk modulus $B$, and (iii)  conclusively demonstrate that surface migration of oligomers in an end-linked polymer gel is suppressed in comparison to a polymer-oligomer mixture} using (i) coarse grained molecular dynamics (CGMD) and (ii) mesoscale hydrodynamics (CHC) simulations (See $SI$ for details). 

\textit{Methods:} We perform CGMD simulations of $10:90$ (\textit{i.e.} $10\%$ oligomer) $OP$ and $OG$ systems using a Kremer-Grest type bead spring model\cite{p:kremer.jcp.v92.p5057.y1990,p:cdgupta.jcp.v140.p244906.y2014} using GROMACS\cite{GROMACS}. The gel matrix of the OG system is constructed by permanently cross-linking terminal beads of two polymers that lie within a distance $\zeta \approx R_{0}$, where $R_{0}$ is the bond distance, (see SI for details). The mesh size of such a system is tuned by changing the relative volume fraction of polymers that make up the network. Interaction strengths among $A$ and $B$ polymers are chosen such that $\epsilon_{AA}=\epsilon_{BB}=2 \epsilon_{AB}=\epsilon$. The system is equilibrated in a box following a temperature quench with periodic boundary conditions along $x$ and $y$ directions and two walls placed at $z=0$ and $z=d$. The wall at $z=0$ preferentially attracts the oligomers (modeling differing surface free energies of oligomers) while the wall at $z=d$ is neutral to both species (see \textit{SI}). 

Configuration snapshots of $OG$ system undergoing simultaneous phase separation and surface migration, obtained from CGMD simulations at different scales of resolution are shown in Fig.\ref{fig01}. The system is quenched from a high temperature $T_{i}=10$ to $T_{f}=1$ (in dimensionless units) and evolved for $\tau = \tau_{LJ} \times 10^{5}$ to ensure thermodynamic equilibrium. Compared to the $OP$ system (see \textit{SI} movies) the phase-separation and thereby surface migration process in arrested in gels. This is evidenced by the presence of \textit{(a)} more oligomer droplets that are smaller in size in comparison to $OP$ systems, \textit{(b)} thinner wetting layer, and \textit{(c)} a narrower depletion region just below the wetting layer (see final configuration of oligomers in \textit{SI} movies). The arrested coarsening and migration behavior is seen in panels (b) and (c) where the oligomer droplets are stuck in a cage formed by end linked polymers. 

Since phase separation is inherently a ``slow'' phenomena, it is difficult to faithfully model it for experimental time scales using detailed CGMD simulations. Meso-scale simulations however access much larger length-scales and longer time-scales. We therefore augment our CGMD simulations with a mesoscale model of phase separation dynamics with the Flory-Huggins free energy functional describing the bulk thermodynamics and local potentials mimicking the preferential surface affinity of oligomers. As the oligomers do not evaporate out of the system, the number of polymers and oligomers in our system is conserved. We therefore employ a time-dependent Ginzburg-Landau formalism using model $B$ dynamics that takes into account the conserved nature of the order parameter\cite{p:cahnhilliard.jcp.v28.p258.y1958,p:hohenberghalperin.rmp.v49.p435.y1977}. We solve the non-linear diffusion equation for the order-parameter field, \textit{i.e.} dynamic oligomer concentration profiles with appropriate boundary conditions to match against similar data obtained from CGMD simulations.

The dynamic concentration profiles of oligomers  $\phi({\bf r}, t)$ for a polymer-oligomer and gel-oligomer system satisfies 
\begin{equation}
\frac{\partial \phi({\bf r},t)}{\partial t} = \nabla \cdot \left[ M \nabla \frac{\delta F[\phi({\bf r},t)]}{\delta \phi({\bf r},t)} + \theta({\bf r}, t) \right], \label{e:diff_eqn}
\end{equation}
where $M$ is the mobility, assumed to be composition independent and the local chemical potential $\mu(\phi({\bf r}, t)) = \frac{\delta F[\phi({\bf r},t)]}{\delta \phi({\bf r},t)}$. An additive vectorial  conserved noise $\theta({\bf r}, t)$ in Eq.\ref{e:diff_eqn} modelling solvent effects, satisfying $\langle \theta_{i}({\bf r}, t) \rangle$ = 0, and $\langle \theta_{i}({\bf r}, t) \theta_{j}({\bf r^{\prime}}, t^{\prime}) \rangle = 2 M k_{B} T \delta_{ij} \delta ({\bf r} - {\bf r^{\prime}}) \delta (t - t^{\prime})$ ensures thermodynamic equilibrium at long times. Since equilibrium bulk concentration of polymers in our system are far from the symmetry point $\phi_{\infty} = 1/2$ where domain coarsening proceeds via spinodal decomposition, we do not include explicit hydrodynamics interactions in our meso-scale model\cite{p:tanaka.prl.v71.p3158.y1993,p:tanaka.pre.v72.p041501.y1995}.

The free energy functional for an incompressible binary fluid mixture, in two space dimensions, confined between selectively attracting walls (surfaces), located at $z=0$ and $z=d$ is given by 
\begin{eqnarray}
F[\phi({\bf r})]/k_{B}T = \frac{1}{a^2} \int_{0}^{d} \int_{0}^d [f_{FH}(\phi) &+& k(\phi) (\nabla \phi)^2 + f_0(\phi)\delta(z) \nonumber \\ &+& f_d(\phi)\delta(z-d) ] dx dz, \label{e:FH_surf_func}
\end{eqnarray}
where $F$ is the free-energy, and $z$ and $x$ are the coordinates perpendicular and parallel to the wall, respectively, and $a$ is the Flory-Huggins lattice spacing. The first term in Eq.~\ref{e:FH_surf_func} is the bulk free energy and the second term accounts for energy costs associated with the spatial gradients of the composition field with a stiffness coefficient $k(\phi) = \frac{a^2}{36 \phi (1 - \phi)}$. Note that surface free-energies $f_0(\phi_{1})$ and $f_d(\phi_{D})$ have dimensions of length such that $F[\phi({\bf r})]$ is dimensionless. The functional forms of $f_{0}(\phi_1) = h_{0} \phi_{1} + \frac{1}{2} g_{0} \phi_{1}^{2}$ and $f_{d}(\phi_{D}) = h_{D} \phi_{D} + \frac{1}{2} g_{D} \phi_{D}^{2}$, respectively. As in our CGMD simulations the wall at $z=0$ attracts the oligomer $B$ while the wall at $z=d$ is neutral to both the components. We study the approach to equilibrium, when this mixture is quenched to the two phase region, starting from an initial uniform phase, which is thermodynamically unstable, for a $OP$ and $OG$ mixture, with the component $A$ (having local composition $\phi({\bf r}, t))$ denoting the polymer with degree of polymerisation, $N_{A}$ or the gel, and an oligomer $B$ (with local composition $(1 - \phi({\bf r}, t))$), with degree of polymerisation, $N_{B}$, respectively. 
 
The dimensionless Flory-Huggins free energy for a polymer-oligomer mixture is given by, Eq.~\ref{e:FH_polym_olig}, 
\begin{equation}
f_{FH}(\phi) = \frac{\phi}{N_A} \ln(\phi) + \frac{(1 - \phi)}{N_B} \ln(1 - \phi) + \chi \phi (1 - \phi), \label{e:FH_polym_olig}
\end{equation}
and the Flory-Rehner free energy describing the gel-oligomer mixture is given by, Eq.~\ref{e:FH_gel_olig},
\begin{eqnarray}
f_{FHE}(\phi) &=& \frac{(1 - \phi)}{N_B} \ln(1 - \phi) + \chi \phi (1 - \phi) \nonumber \\ &+& B(\frac{\phi_{\infty}}{2})
[(\frac{\phi}{\phi_{\infty}})^{2/3} + 2(\frac{\phi}{\phi_{\infty}})^{1/3} -3], \label{e:FH_gel_olig}
\end{eqnarray}
where $\chi$ is the Flory-Huggins interaction parameter. The bulk concentration of the polymers that make up the gel is denoted by $\phi_{\infty}$, which is identified here as the region in the vicinity of $z=d$, and $B$ denotes the bulk modulus of the gel. We numerically integrate Eq.~\ref{e:diff_eqn} for both forms of the free energies Eq.~\ref{e:FH_polym_olig} and Eq.~\ref{e:FH_gel_olig} with the boundary conditions described earlier (see \textit{SI} for details). For bulk simulations we impose a periodic boundary condition along all directions, while in the presence of walls we impose a zero flux boundary condition at the walls and periodic boundary condition along the lateral dimensions. This ensures that the order parameter is conserved throughout the evolution process. The initial $\phi(\bf{r}, t)$ field configuration for mesoscale simulations on a ($L \times L$) lattice with $L=50$ is chosen to be $\phi(\bf{r}, 0) = \phi_{\infty} + \delta \phi$, with $\phi_{\infty}$ being the equilibrium bulk concentration of a polymer/gel and $\delta \phi$ is a uniformly distributed random number in the interval $\left[-0.05, 0.05 \right]$. \\

We coarse-grain particulate configuration snapshots of $CGMD$ simulations\cite{p:sdas.pre.v65.p026141.y2002} to obtain oligomer concentration $\phi({\bf r}, t)$ to compare against $MH$ results following a majority rule. The simulation box is divided into cubes of size $\sigma \approx b$, where $b$ is the bead diameter and the number of monomers belonging to polymer $n_{A}$ and oligomer $n_{B}$ counted. The coarse-grained order parameter field for the $i$-th cell $\phi_{i} = +1$ if $n_{A} > n_{B}$, otherwise  $\phi_{i} = -1$. For the simulations in presence of walls, the coarse-grained $\phi_{i}$'s, the one-dimensional density of oligomers, as a function of the depth from the upper wall, is obtained by performing an average over the two lateral dimensions. The equal time spatial correlation function in bulk mixtures

\begin{eqnarray}
C(r, \tau) = \langle \phi(0,\tau)\phi(r,\tau) \rangle - \langle \phi(0,\tau) \rangle \langle \phi(r,\tau) \rangle, \label{e:corr_fun}
\end{eqnarray}
where $r$ is the radial distance between the centres of two cubes, and $\tau$ is the time elapsed since the instantaneous quench, is used to study the phase-separation dynamics. The angular brackets in Eq.\ref{e:corr_fun} indicate averaging over independent initial configurations and the first zero crossing of $C(r, \tau)$ determines the domain size $\ell(\tau)$. 

\begin{figure}[t]
\includegraphics[width=4.25cm]{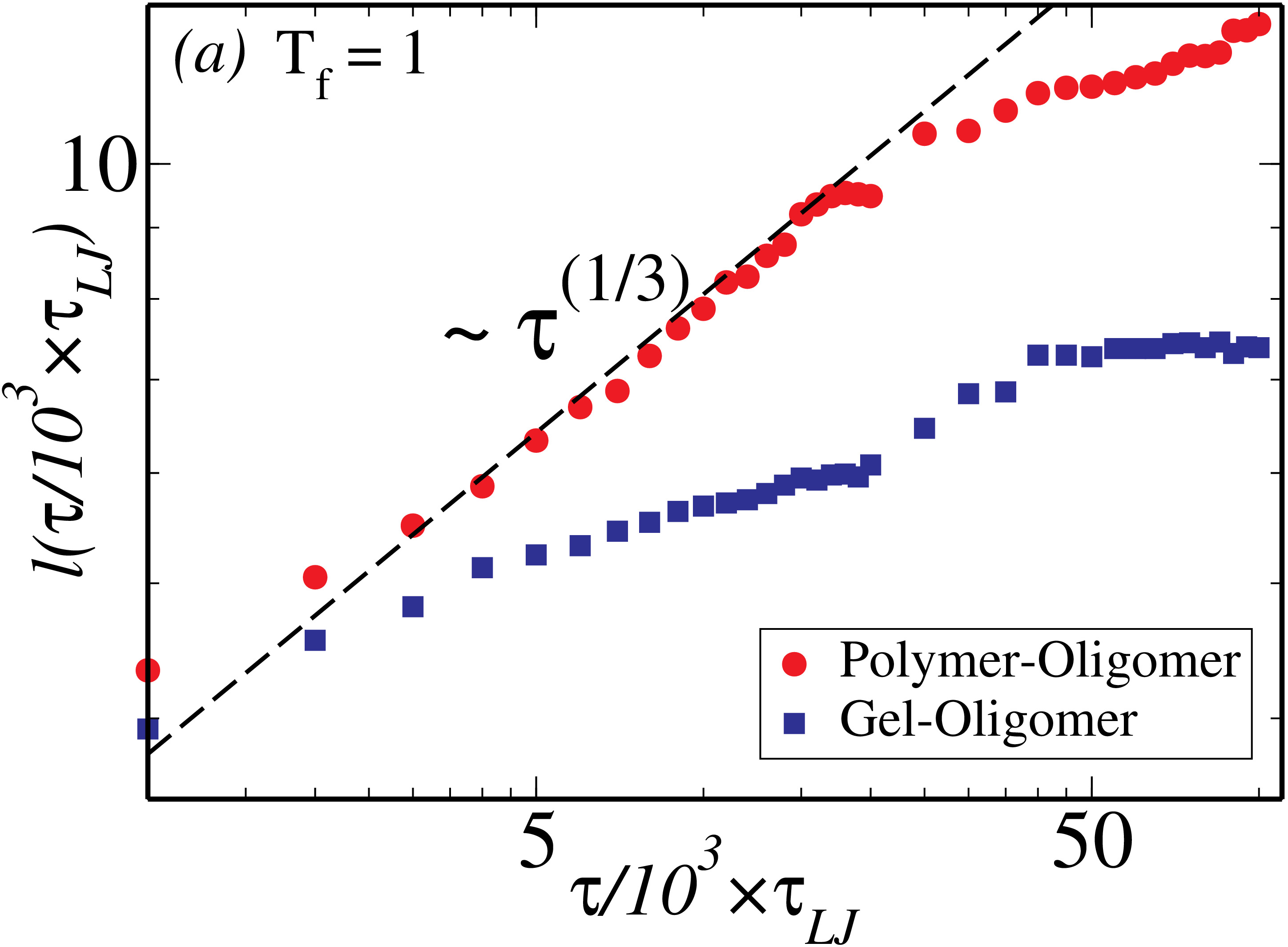}
\includegraphics[width=4.25cm]{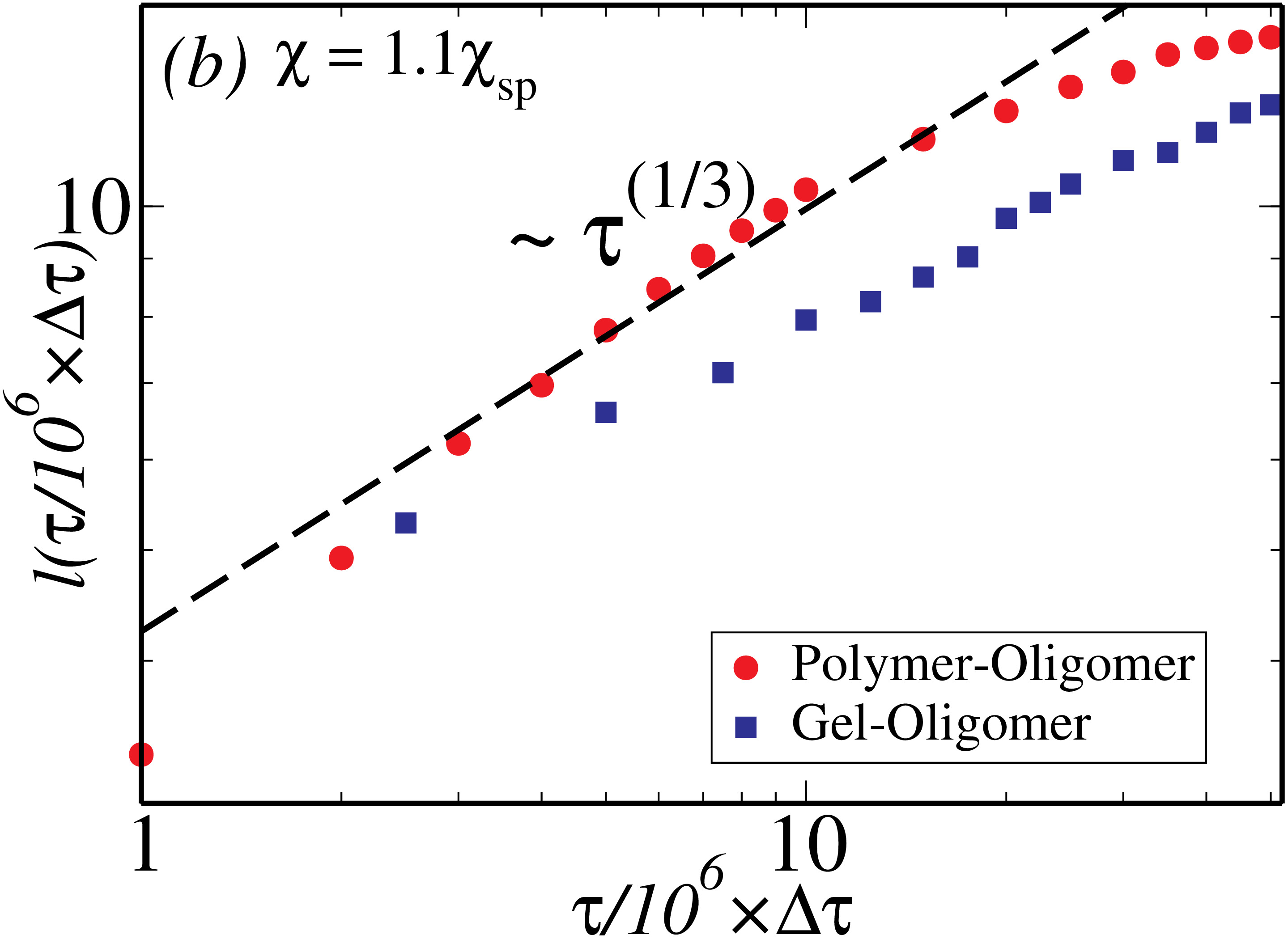}\\
%\end{tabular}
\caption{The time-dependence of the coarsening length as a function of time computed from bulk simulations for polymer-oligomer mixture (shown in red filled circles) and gel-oligomer mixtures (shown in blue-filled squares) simulated via MD simulations (panel (a)) and via meso-scale simulations, where the gel has a bulk modulus, $B = 0.05$, (panel (b)).}
\label{fig02}
\end{figure}

\textit{Results:} The time-dependence of the coarsening length-scale is shown in Fig.\ref{fig02}, with panel (a) and (b) showing results from bulk MD simulations and bulk mesoscale simulations respectively, with filled circles denoting coarsening in polymer-oligomer mixture and filled squared denoting coarsening in a gel-oligomer mixture with bulk modulus, $B = 0.05$. In both cases the domain size grows as a function of time as $\ell(\tau) \sim \tau^{1/3}$ following a Lifshitz-Slyozov law. The phase separation process is arrested in gels showing $\ell(\tau)$ saturating at higher values of $\tau$. The saturation increases with increasing bulk modulus. \\   
%%%%%%%%%%%%%%%%
\begin{figure}[t]
\includegraphics[width=8cm]{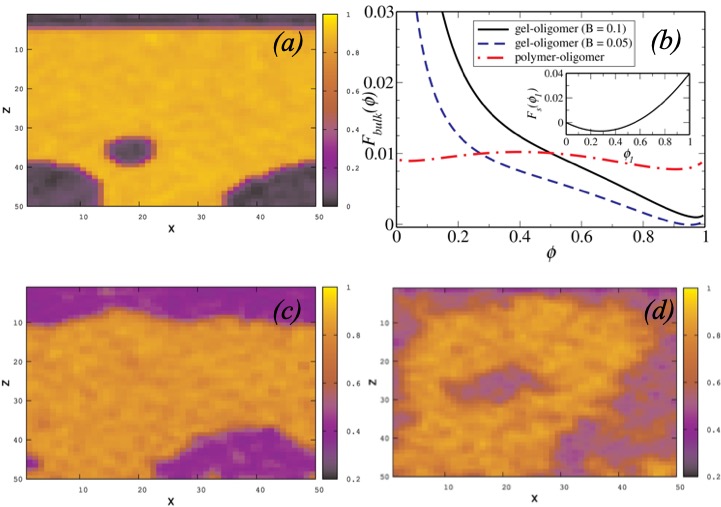} 
\caption{Oligomer concentration configurations for $OP$ ((a)) and $OG$ ((c) and (d)) systems showing polymer rich (yellow) and oligomer rich (dark) domains obtained from mesoscale simulations with oligomers wetting the wall at $z=0$. Panel (b) shows Flory-Huggins (red dashes) and Flory-Rehner (blue dash-dotted line $B=0.05$, and black solid line $B=0.1$) forms of bulk free energies corresponding to these mixtures. Inset (b) shows variation of surface free energy of oligomers for interaction parameters $g_0 = $ and $h_0 = $ as a function of the surface oligomer concentration $\phi_{1}$.\label{fig03}}
\end{figure}
%%%%%%%%%%%%%%%%
In order to match our results with CGMD simulations we choose $\phi_{\infty} = 0.7$ which corresponds to a $30:70$ asymmetric mixture of oligomers and polymers/gel having polymerisation index $N_{A} = 100$, and $N_{B} = 50$ respectively. We set $\chi = 1.1 \chi_{sp}$, where $\chi_{sp}$ corresponds to the critical value of the Flory parameter for the above parameters and $\phi_{\infty}=0.7$. Two values of bulk modulii $B=0.05$ and $B=0.1$ have been used and the computed thermodynamic quantities are averaged over $N_{r} = 10$ different initial configurations. We numerically integrate the non-dimensionalised version of Eq.\ref{e:diff_eqn} accounting for the conserved noise following a forward Euler scheme (see \textit{SI}). 

For the top surface $z=0$ preferentially attracting oligomers a wetting transition is observed. The migrant concentration configurations close to equilibrium having a small chemical potential gradient $\delta \mu \approx 0$, obtained by numerically integrating Eq.\ref{e:diff_eqn} for long times $t \rightarrow \infty$ for both systems are shown in Fig.\ref{fig02}. At long times, the phase separation is nearly complete for $OP$ systems resulting in the formation of a thick wetting layer. In contrast, the coarsening process is arrested in gels resulting in a diffuse thin wetting layer that decreases monotonically on increasing the bulk modulus, panel (c), (d). These results can be understood from the variation of the bulk free energy as a function of the oligomer concentration $\phi$ for both systems. In the absence of elastic interactions the system admits two minima corresponding to an equilibrium phase that is $A/B$ rich.  For a gel, elastic interactions result in lowering the free energy of an oligomer rich state. If the surface affinity of the oligomers (set by $g_0$, $h_0$, $g_D$ and $h_D$) is insufficient to overcome the cost of elastically deforming a polymeric cage that traps the oligomer droplets, the equilibrium state is one with oligomers inside the gel resulting in a thinner diffuse wetting layer. 
%%%%%%%%%%%%%%%%%%%%%%%
\begin{figure}[t]
\centering
\includegraphics[width=8cm]{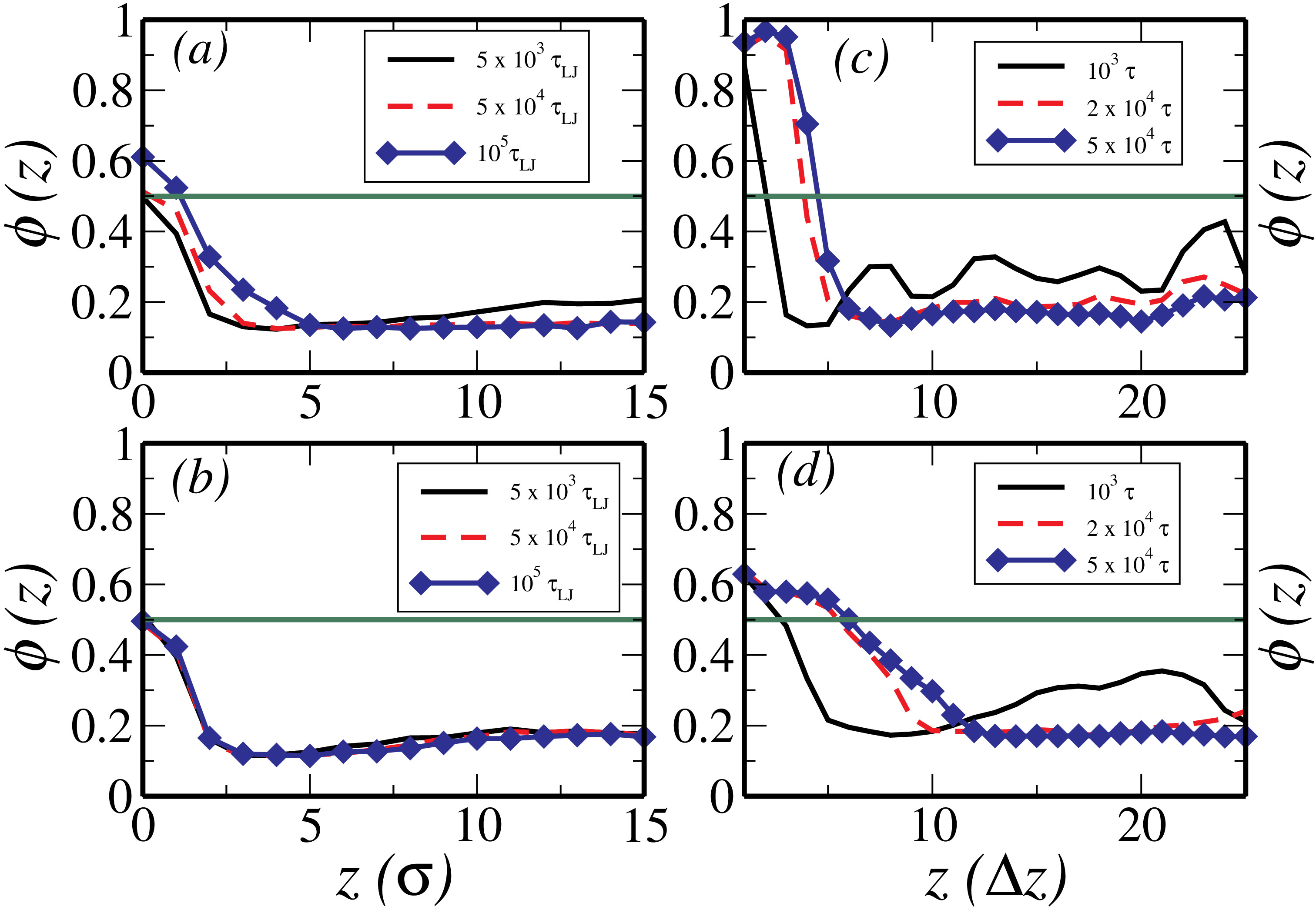} 
\caption{The time-evolution of the oligomer density computed from MD simulation in a polymer-oligomer system (panel (a)) and a gel-oligomer system (panel(b)) 
following a quench from an initial temperature of $T_{i} = 10$ to a final temperature of $T_{f} = 1$. Panels (c) and (d) shows the time evolution of the oligomer density computed from meso-scale simulations, following a quench to the two-phase region, for a polymer-oligomer and a gel-oligomer mixtures, respectively.}
\label{fig:olig_dens_evolution}
\end{figure} 
%%%%%%%%%%%%%%%%%%%%%%%

Fig.\ref{fig:olig_dens_evolution} shows the time evolution of oligomer concentration with an attractive surface. Panels (a) and (b) shows profiles obtained for a fraction $\phi=0.9$ of polymer and gel molecules respectively following an instantaneous quench from an initial temperature $T_{i}=10$ to $T_{f}=1$. The migrant density in the vicinity of the upper wall at $z=0$ is $\phi(z) >0.5$ for the $OP$ mixture in panel (a), whereas,  $\phi(z)<0.5$ for the $OG$ system in panel (b). Similarly (c) and (d) of Fig.\ref{fig:olig_dens_evolution} shows density profiles obtained from mesoscale simulations for parameters ($\phi_{\infty}=0.7$, and $\chi_{sim}=1.1 \chi_{sp}$) as described above. The characteristic time and space discretisation scales of the mesoscale simulations are dependent on the thermodynamic state point (see \textit{SI}). When comparing results from different simulations we have rescaled the raw data such that all temporal and spatial scales in Fig.\ref{fig:olig_dens_evolution} are equal. \\
%%%%%%%%%%%%%%%%%%%%%%%%%%%%
\begin{figure}[htpb]
\centering
\begin{tabular}{cc}
\includegraphics[width=4cm]{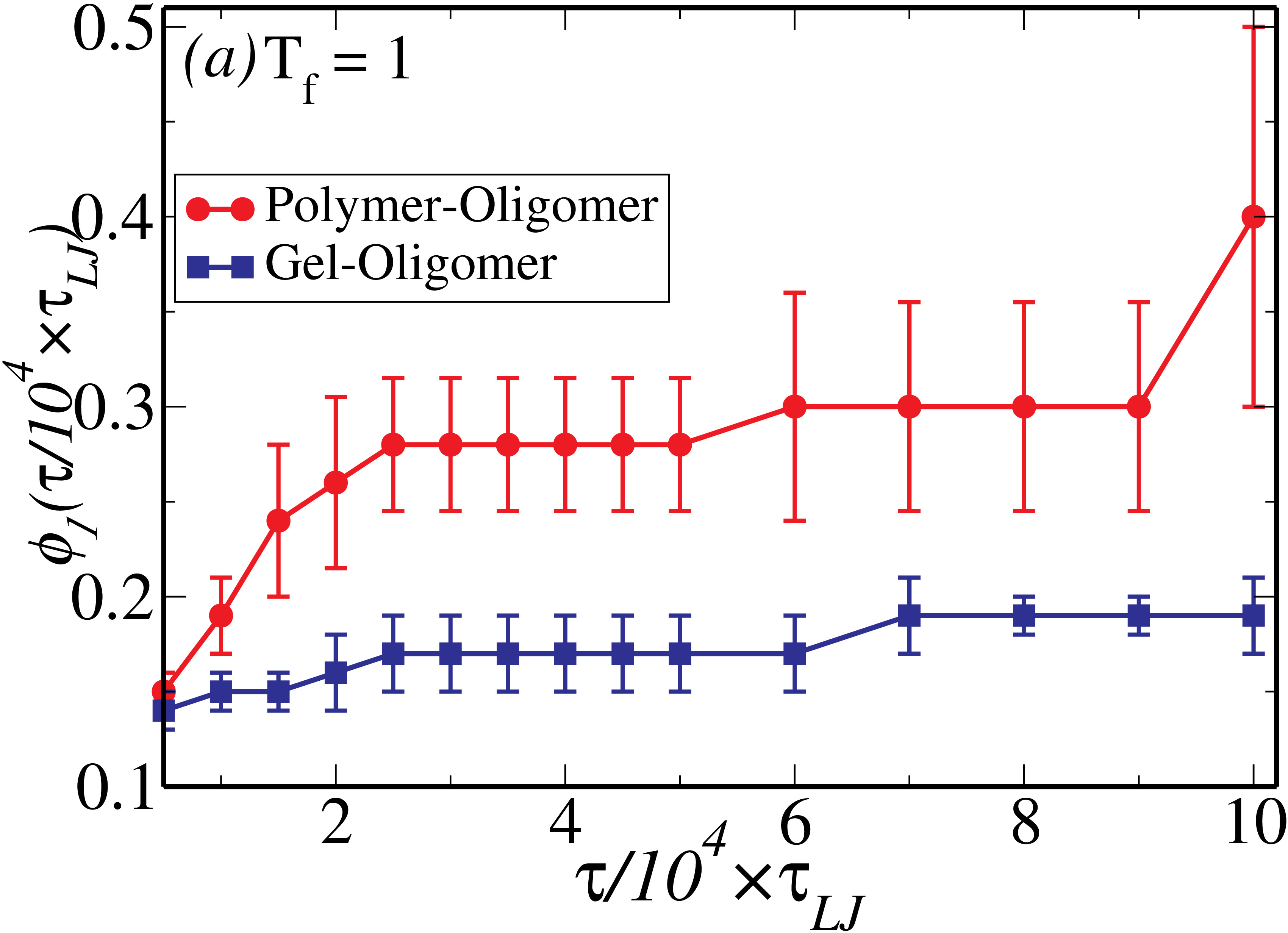}&
\includegraphics[width=4cm]{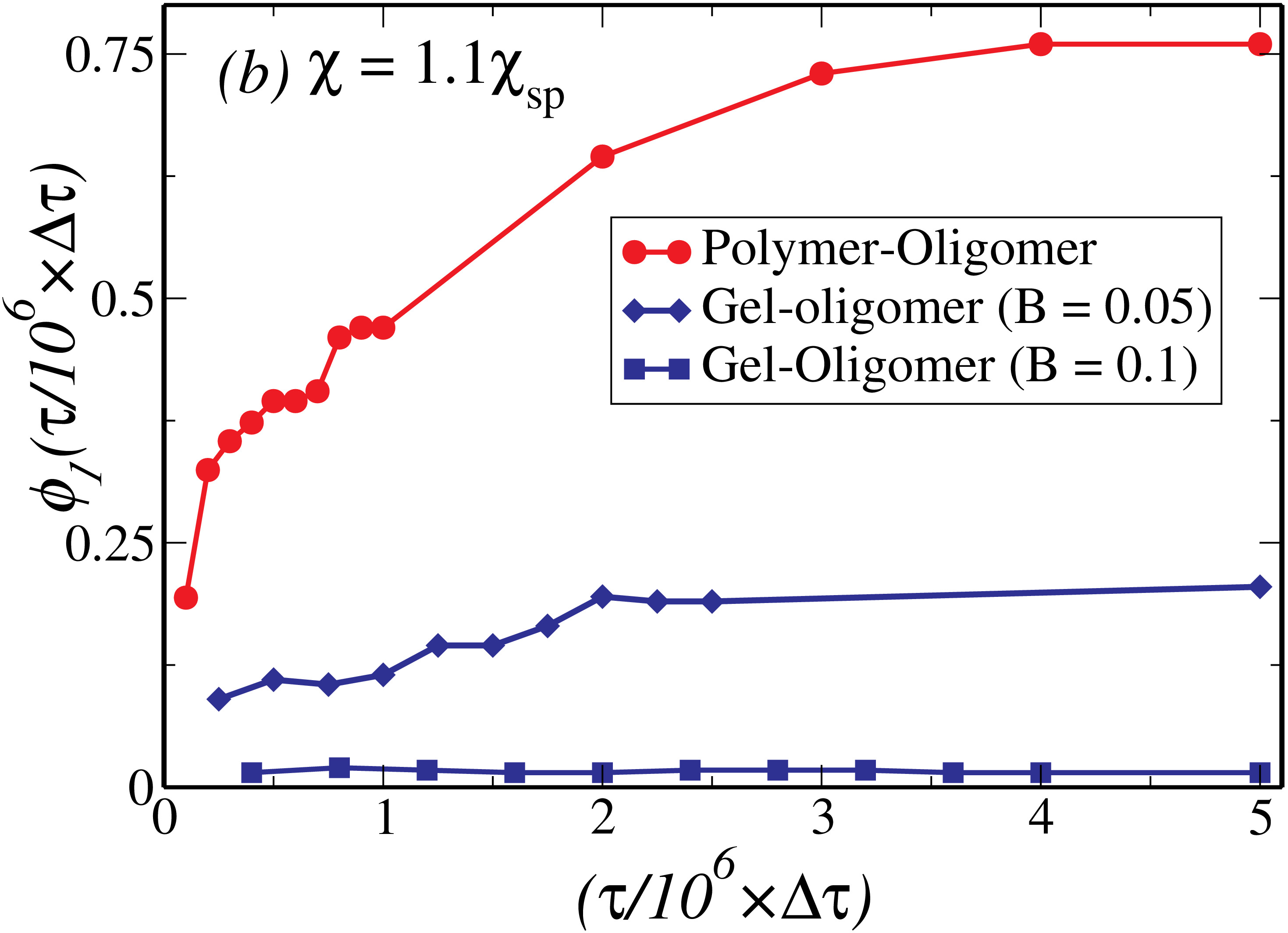}\\
\end{tabular}
 \caption{The fraction of migrant molecule computed from an MD simulation of a polymer-oligomer mixture and a gel-oligomer mixture is shown in panel (a), while panel (b) shows the same from a meso-scale mixture of a polymers and oligomers and a mixture of gel and oligomers for two values of gel bulk moduli.}
    \label{fig:migrant_frac}
\end{figure}

The oligomer density profiles shown in Fig.\ref{fig:olig_dens_evolution} is used to compute the migrant fraction at the attractive surface as a function of time. In CGMD simulations this is computed by counting the number of particles between $z=0$ and the first minimum of the density profile $\phi(\bf{r}, t)$ at $z \approx 4 \sigma$. In mesoscale simulations the migrant fraction $\phi_{1} = \int^{\ell_{w}}_{0} \left( \phi(z) - 0.5 \right) d z$, \textit{i.e.} the area of $\phi(z)$ above the line $\phi(z)=0.5$. Fig.\ref{fig:migrant_frac} (a) and (b) shows the time variation of the migrant fraction for the $OP$ and $OG$ systems by these methods. As evidenced in the oligomer density profiles, increasing the bulk modulus causes a decrease in the migrant fraction. A similar dramatic slow growth of the wetting layer thickness as a function of increasing bulk modulus is observed in CGMD simulations (see \textit{SI} Fig.4). 

%The horizontal lines in all the panels allows us to compute the thickness of the wetting layer, which we define as the depth at which $\psi$(z) crosses 0.5. 
\textit{Discussion:} In conclusion we have showed that surface migration in $OG$ systems can be significantly reduced by increasing the elastic modulus of the gel. Our CGMD simulations show that for these systems oligomer droplets are stuck in the gel meshwork, leading to a phase-separation arrest that modifies both the domain growth law and surface segregation kinetics. The phase separation proceeds either via nucleation, \textit{i.e.} growth and coalescence of droplets or spinodal decomposition \textit{i.e.} unstable growth. The nucleation and growth phenomena is driven by surface tension and mass diffusion among different sized droplets and leads to the $LSW$ domain growth law $\ell(\tau) \sim \tau^{1/3}$ for systems with Ising symmetry as seen from simulations\cite{p:bray.advphys.v43.p357.y1994} and asymptotic analysis\cite{p:lifshitz.jpcsolid.v19.p35.y1961,p:wagner.zelectrochem.v65.p581.y1961}. Such a growth law has been explored for binary \cite{p:Petscheck.jcp.v79.p3443.y1983} and multicomponent fluids\cite{p:sdas.pre.v65.p026141.y2002}, and polymer mixtures\cite{p:binder.jcp.v79.p6387.y1983,p:muthu.prl.v63.p2072.y1989,p:muthu.jcp.v92.p6899.y1990,p:holyst.jcp.v117.p1886.y2002,p:nigelclake.macromol.v37.p1952.y2004,p:cdgupta.jcp.v140.p244906.y2014} and has been verified in experiments\cite{p:carlow.prl.v78.p4601.y1997,p:voorhees.prl.v82.p2725.y1999} (see \cite{p:bray.advphys.v43.p357.y1994} for a review). End linking polymers result in effectively slowing down relaxation mechanisms leading to a ``dynamical asymmetry'' among the constituents leading to a dramatic slow down of the LSW law. The system thus exhibits characteristics of viscoelastic phase separation\cite{p:tanaka.prl.v71.p3158.y1993,p:tanaka.pre.v72.p041501.y1995}. While size disparity in soft material mixtures leading to formation of transient networks\cite{p:pratibha.softmat.v13.p2330.y2017} is common, it is fundamentally different from our case where the network structure is permanent. Departures from the LSW kinetics in bulk fluids can also occur in systems where phase separation is coupled to chemical reactions. A coupled gelation, phase-separation and surface migration study with competing time-scales that lead to novel phenomena will be reported in a future study\cite{p:self.future.vxx.pxx.y2019}.

The kinetics of surface-directed spinodal decomposition (SDSD) is a rich non-equilibrium, many-body phenomena where the dynamic effects of surface wetting and bulk phase separation are coupled in a non-trivial fashion \cite{p:puri.pra.v46.pR4487.y1992,p:marko.pre.v48.p2861.y1993,p:achakrabarti.pre.v46.p4829.y1992,p:puri.pre.v49.p5359.y1994,p:puri.prl.v86.p1797.y2001,p:puri.pre.v66.p061602.y2002}. An early time surface interaction specific growth law that leads to late time LSW kinetics is observed for the minority component being attracted by the surface\cite{p:puri.pre.v66.p061602.y2002}. Experiments for SDSD for $OP$ systems show that the wetting layer thickness grows logarthmically as a function of time for a shallow quenche and follows\cite{p:goegehan.pre.v62.p940.y2000,p:geogeghan.progpolysci.v28.p261.y2003} LSW growth for a deep quench. Several factors, \textit{e.g.} surface adsorption\cite{p:sdas.prl.v96.p016107.y2006,p:binder.jstatphys.v138.p51.y2010}, surface roughness\cite{p:scoveney.prl.v113.p218301.y2014,p:nigelclarke.prl.v111.p125701.y2013}, and confinement, can modify surface migration kinetics in polymer mixtures leading to novel phenomena, \textit{e.g.} lateral phase separation\cite{p:nigelclarke.prl.v111.p125701.y2013,p:nigelclarke.pre.v89.p062603.y2014}.

The generic multiscale framework developed in this paper is suitable for rational design of oligomer-polymer/gel mixtures with predictable surface migration and phase behaviour.  Our non-equilibrium phase ordering kinetics presented here can also be extended to the biological domain via the incorporation of an active term. In particular, we believe that the techniques and results presented here are directly applicable to viscous cellular environments where membraneless organelles sort cellular components via liquid-liquid phase separation\cite{p:brangwynne.sci.v324.p1729.y2009,p:dufresne.prx.v8.p011028.y2018}. 
 
\begin{acknowledgments}
%\label{acknow}
\textit{Acknowledgements:} BC and BM acknowledge funding support from EPSRC via grant $EP/P07864/1$, and P\&G, Akzo-Nobel, and Mondelez Intl. Plc. 
\end{acknowledgments}

%\bibliographystyle{apsrev4-1}
%\bibliography{references}

\end{document}